\documentclass[conference]{IEEEtran}
\IEEEoverridecommandlockouts
\usepackage{cite}
\usepackage{amsmath,amssymb,amsfonts}
\usepackage{graphicx}
\usepackage{textcomp}
\usepackage{xcolor}
\usepackage{booktabs}
\usepackage{multirow}
\usepackage{caption}
\usepackage{algorithm}
\usepackage{algpseudocode}
\usepackage{commath}
\usepackage{url}
\usepackage{booktabs}
\usepackage{ctable}

\begin{document}
\title{
\huge Distributed Software-Defined Network Architecture for Smart Grid Resilience to Denial-of-Service Attacks} 

\author{\IEEEauthorblockN{\hspace{0.08in}Dennis Agnew$^{1}$ \hspace{0.08in} Sharon Boamah$^{1}$ \hspace{0.08in} Reynold Mathieu$^{1}$\hspace{0.08in} Austin Cooper$^{1}$ 
\hspace{0.08in}Janise McNair$^{1}$ \hspace{0.08in} Arturo Bretas$^{1,2}$}
\IEEEauthorblockA{\textit{Department of Electrical and Computer Engineering},
\textit{University of Florida, Gainesville, FL$^{1}$} \\
\textit{Distributed Systems Group, 
Pacific Northwest National Laboratory, Richland, WA$^{2}$}\\
\{dennisagnew, sharonboamah, reynold.mathieu, austin.cooper\}@ufl.edu, \{mcnair, arturo\}@ece.ufl.edu}\vspace*{-0.75cm}
\thanks{This work was partially funded by the U.S. Department of Energy under Contract DE-AC05-76RL01830.}
}

\maketitle

\begin{abstract}
An important challenge for smart grid security is designing a secure and robust smart grid communications architecture to protect against cyber-threats, such as Denial-of-Service (DoS) attacks, that can adversely impact the operation of the power grid. Researchers have proposed using  Software Defined Network 
frameworks to  enhance cybersecurity of the smart 
grid, but there is a lack of benchmarking and comparative analyses among the many techniques.
In this work, a distributed three-controller software-defined networking (D3-SDN) architecture, benchmarking and comparative analysis with other techniques is presented.  The selected distributed flat SDN architecture divides the network horizontally into multiple areas or clusters, where each cluster is handled by a single Open Network Operating System (ONOS)  controller.
A case study using the IEEE 118-bus system is provided to compare the performance of the presented ONOS-managed D3-SDN, against the POX controller. In addition, the proposed  architecture outperforms a single SDN controller framework by a tenfold increase in throughput; a reduction in latency of $>20\%$; and an increase in throughput of approximately $11\%$ during the DoS attack scenarios.
\end{abstract}


\section{Introduction}
Cyber-related smart grid (SG) system security has been a growing concern over the last decade.
The current focus of research for power grid cybersecurity is operation technology (OT) and information technology (IT) systems reliability, whereas the development of an interdependent SDN management framework for SGs communications is still emerging.
Traditional power systems are often protected by isolated and uncoordinated equipment that provides ad hoc solutions to each protection challenge. Other solutions are based in ongoing research for power system state estimation \cite{monticelli1999state}. Although these techniques can be effective, these technologies are not integrated to communicate online with one another, and thus they are vulnerable to distributed cyber-attacks,  denial-of-service (DoS), distributed denial-of-service (DDoS), man-in-the-middle (MITM), and false data injection (FDI)  \cite{Allen2022Starke}.
These types of attacks can influence data from several layers of the grid's physical structure and ultimately disrupt system service.
A unified methodology to recognize multiple attacks conducted from various layers inside the SG is still to be realized. 

In our previous studies \cite{Allen2022Starke,agnew2022implementation,aljohanicross}, we have explored the use of cross-layered data from both the communication and power grid layer for a Cross-Layer Ensemble CorrDet with Adaptive Statistics (CECD-AS) model to identify and characterize various cyber attacks. We have also introduced software defined networks (SDNs) as a possible underlying architecture to facilitate SG cybersecurity~\cite{agnew2022implementation,starke2018toward}.

This work is a continuation of our previous study \cite{agnew2022implementation}, and provides a statistical analysis of a distributed flat topology for the SDN. We further present a comparison with the common alternative controller, the POX controller~\cite{kaur2014network,cokic2019software, mahmood2021s}. 
Several studies such as~\cite{ghosh2017security, qureshi2020distributed, hussain2019model} have suggested a distributed SDN architecture for smart grids, but none provide a comparison of performance between distributed SDN framework and the single controller approach used in the POX controller-based literature. This work seeks to provide such a comparative analysis, and makes the following specific contributions towards the state-of-the-art:
\begin{enumerate}
    \item A benchmark study that compares and analyzes distributed versus centralized SDN  management layer solutions for Smart Grids.
    \item A benchmark study for Smart Grid SDN frameworks to evaluate throughput resiliency of distributed vs centralized SDN frameworks against DoS attacks. 
\end{enumerate}

The remainder of the paper is organized as follows. Section~\ref{Background} presents background information on the SDN framework and the network statistics used, while Section~\ref{Implementation_of_SDN} contains data flow information of the framework, and provides the implementation aspects. 
A case study is presented in Section~\ref{Case Study}, followed by concluding remarks in Section~\ref{Conclusion}.

\section{Background}
\label{Background}

\subsection{Software-Defined Networking (SDN)}
\label{SDN_section}
The concept of SDN has become more prevalent for use in network management in recent years, developed by Stanford University to describe Openflow principles and practices~\cite{kreutz2014software}. SDN appeals to networking professionals due to its visibility and network device programmability. SDN improves resource efficiency, network service flexibility, and maintenance costs~\cite{sun2020detecting}. 
One of most popular open-source SDN controller for next-generation SDN and network function virtualization (NFV) is the ONOS controller. The ONOS controller configures and controls networks in real-time without routing and switching control protocols. 
An SDN network design can move the network's routing intelligence to the ONOS controller, improving network administration, response, and visibility to cyber threats.

\begin{figure}[t!]
\centering
\includegraphics[width=.8\linewidth,trim = 0cm 0mm 0cm 0cm,clip]{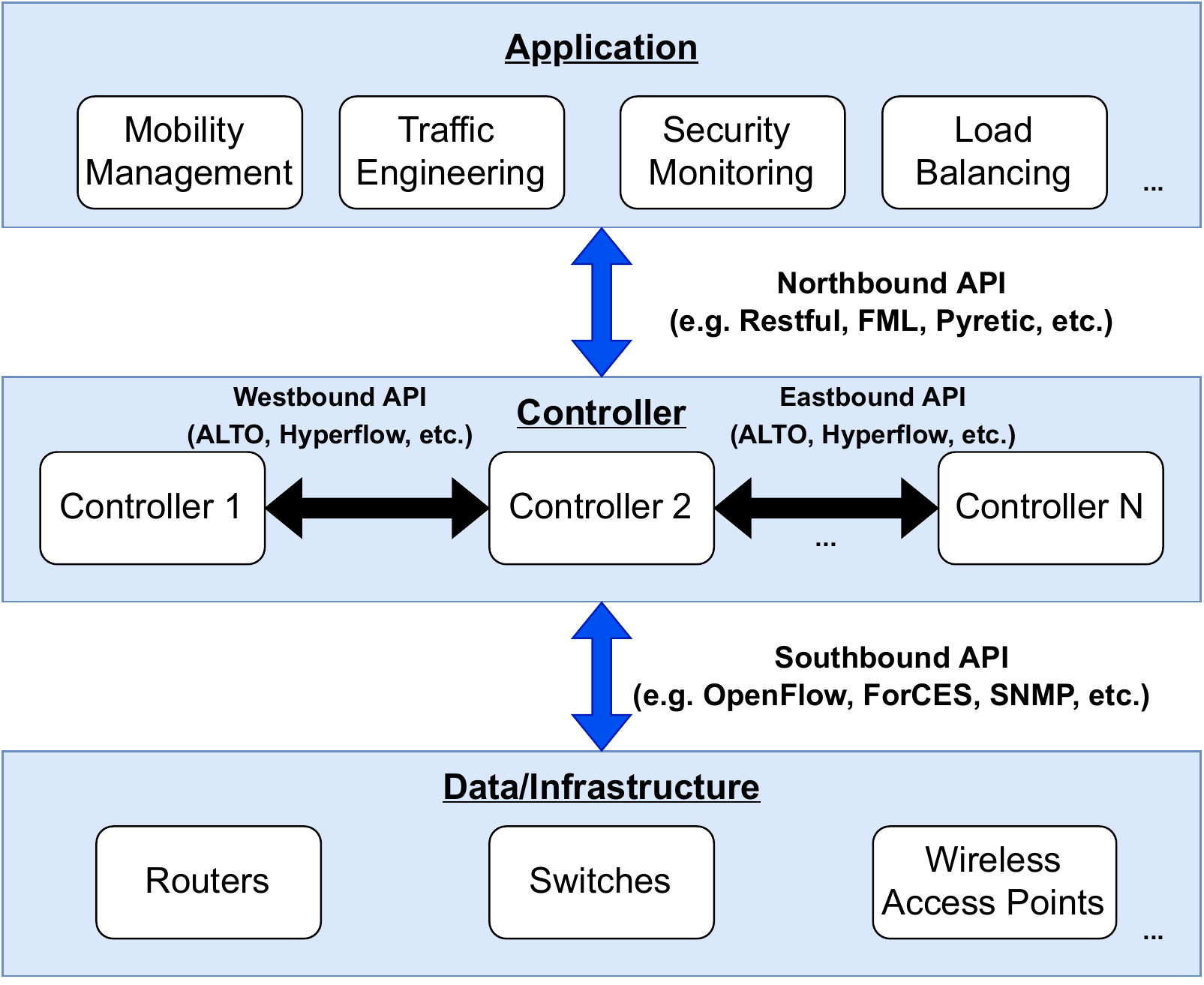}
\caption{General SDN Architecture~\cite{agnew2022implementation}.}
\vspace{-7pt}
\label{fig:SDN_overview}
\end{figure}

Figure~\ref{fig:SDN_overview} depicts a high-level overview of a modern functioning SDN infrastructure. 
 SDN is divided into three planes:
\begin{enumerate}
    \item \textbf{Application Plane:} It includes network administration, policy implementation, and SDN applications for security services.
    \item \textbf{Control Plane:} The network operating system and SDN applications are run by this logically centralized control framework. SDN flows are instructions followed by a packet sequence between the source and destination. Controllers install flows in forwarding device flow tables.
    \item \textbf {Data/Infrastructure Plane:} This layer represents the physical network equipment in the network, a collection of forwarding components that shift traffic flows in response to control plane commands. The infrastructure layer is represented by routers, switches, and access points in the network. 
\end{enumerate}

The SDN architecture planes communicate via application programming interfaces (APIs) to achieve interaction with network control interfaces. The application plane is the topmost layer, affording the network operator use of functional applications to enact policies for energy efficiency, access control, mobility management, and security management. Northbound APIs like FML, Procera, Frenetic, and RESTful connect the application and control layers. These APIs let the network operator communicate with the control layer, so the controller can make infrastructure layer changes. Southbound APIs, e.g., OpenFlow, ForCES, PCEP, NetConf, and IRS link the controller to the data plane, and Westbound and Eastbound APIs, like AlTO or Hyperflow, connect multiple controllers. 

As shown in Figure~\ref{fig:SDN_overview}, controller choice is crucial  for SDN network operation and performance. There are several controllers utilized in SDN literature research, including Python-based OpenFlow controller (POX), Floodlight, OpenDaylight (ODL), RYU, and ONOS, each with its own specialized features and functions. POX, which was based on NOX, the first OpenFlow controller, is a well-known controller that is ideal for rapid prototyping. 
However, the POX controller cannot handle multiple, distributed controllers. The east/westbound API connection, necessary to connect multiple controllers, is not possible. 
On the other hand, due to its simple implementation, it is often the controller of choice for quickly testing SDN frameworks on emulation software like Mininet. 
The POX controller has been used in several Smart Grid studies. 

In this paper, we use an ONOS-based distributed three controller cluster to manage the smart grid SDN network. ONOS  was selected because of its demonstrated performance over its competitors in other applications and its continued development and documentation for real-world applications~\cite{mamushiane2021qos}. 
We use the POX controller as a point of comparison for the distributed framework presented. 

To analyze the performance of SDN, Mininet \cite{github}, an open-source networking software, is used to prototype and emulate SDN networks with hosts, connections, and switches on a single device. %~\cite{kaur2014mininet}.
It uses process-based virtualization and network namespaces in recent Linux kernels to create virtual networks. The Mininet hosts emulate bash processes in a network namespace, so web servers and client software can run normally over Open vSwitch and OpenFlow reference switches. 
Links connect the emulated switches and hosts via Linux kernel virtual ethernet pairs. 

\subsection{Network Performance Statistics}
\label{sec:backgroundstats}
The framework for cross-layered analysis is based on the IEEE 118-bus system, which mimics the Modbus RTU using TCP/IP protocols. 
The communication layer, which resembles the Poisson traffic model \cite{jain1986packet}, transmits packets every four seconds in groups of four. 
Each bus represents the M/M/c queue \cite{haviv2009queues}, i.e. c = 1, with Poisson packet arrival and exponential queue service time. 

The rate of packet arrival is denoted by $\lambda$, while the service rate of packets is represented by $\mu$. Inter-arrival time (IAT) is the difference in time ($\Delta t$) between the arrivals of two or more packets. With parameter $\lambda$, the distribution is exponential. For t $\ge$0, the probability density function is defined as follows:
\begin{equation}
    f(t) =\lambda e ^{-\lambda t}. 
\end{equation}

The average IAT is defined as
\begin{equation}
IAT = \frac{1}{\lambda}.    
\end{equation}
% The service time follows an exponential distribution with parameter $\mu$. The probability density function is as follows:

The service time follows an exponential distribution with service rate $\mu$. The probability density function is defined as follows:
\begin{equation}
    g(s) = \mu e^{-\mu s}, \forall \ge 0
\end{equation}
where $\frac{1}{\mu}$ is the average service time of the system. Utilizing Little's theorem, the total waiting time is represented as transmission delays (TD) as follows:
\begin{equation}
    W = TD = \frac{1}{\mu - \lambda}.
\end{equation}

% The normal distribution of network packet arrivals (i.e., non-attacked packets) into each system was determined by the probability of observing a given number of packet arrivals in a time interval ranging from [0,T]. This equation is utilized to model the bus's traffic volume:
% \begin{equation}
%     P(n\  arrivals\ in\ interval\ T) = \frac{(\lambda T)^{n}e^{-\lambda T}}{n!}
% \end{equation}

% whereas T represents the IAT and $n$ indicates the number of packets. The model for the packet-count (PC) is as follows:
% \begin{equation}
%     PC = \lambda T.
% \end{equation}

Propagation delay (PD) is the time required for a packet to traverse a wire, or link and is defined as follows:
\begin{equation}
\label{PD}
    PD = \frac{d}{s}
\end{equation}
% whereas d equals distance in kilometers and s is the wave propagation speed. Latency (L) is the sum of all possible delays a packet can face during transmission and is denoted as follows:
% \begin{equation}
%     L = D_{TD} + D_{\mu} + D_{Q}+ D_{PD}
% \end{equation}
% whereas $D_{TD}$ is the transmission delay, $D_{\mu}$ is the service time delay, $D_{Q}$ is packet queuing delay, and $D_{PD}$ is the propagation delay.
where $d$ equals distance in kilometers and $s$ is the wave propagation speed in kilometers per second. The total latency ($L$), the sum of all delays a packet faces during transmission, is denoted in this study as follows:
\begin{equation}
    L = D_{TD} + D_{\mu} + D_{PD}
\end{equation}
where $D_{TD}$ is the transmission delay, $D_{\mu}$ is the service time delay and $D_{PD}$ is the propagation delay.
The throughput ($TH$) is defined as the maximum number of bits per second that can be experienced through the communication link or wire. 
$TH$ is defined as follows:
\begin{equation}
    TH \le \frac{RW}{RTT}
\end{equation}
whereas $RW$ is the receive window, or receiver's buffer for incoming data that has not be processed yet, and $RTT$ (round trip time) is the time it takes a packet to travel from a source to a destination and for the acknowledgement to return back.

\section{Distributed SDN Implementation}
\label{Implementation_of_SDN}

\begin{figure}[!h]
\centering
\includegraphics[width=.8\linewidth,trim = 0cm 0mm 0cm 0cm,clip]{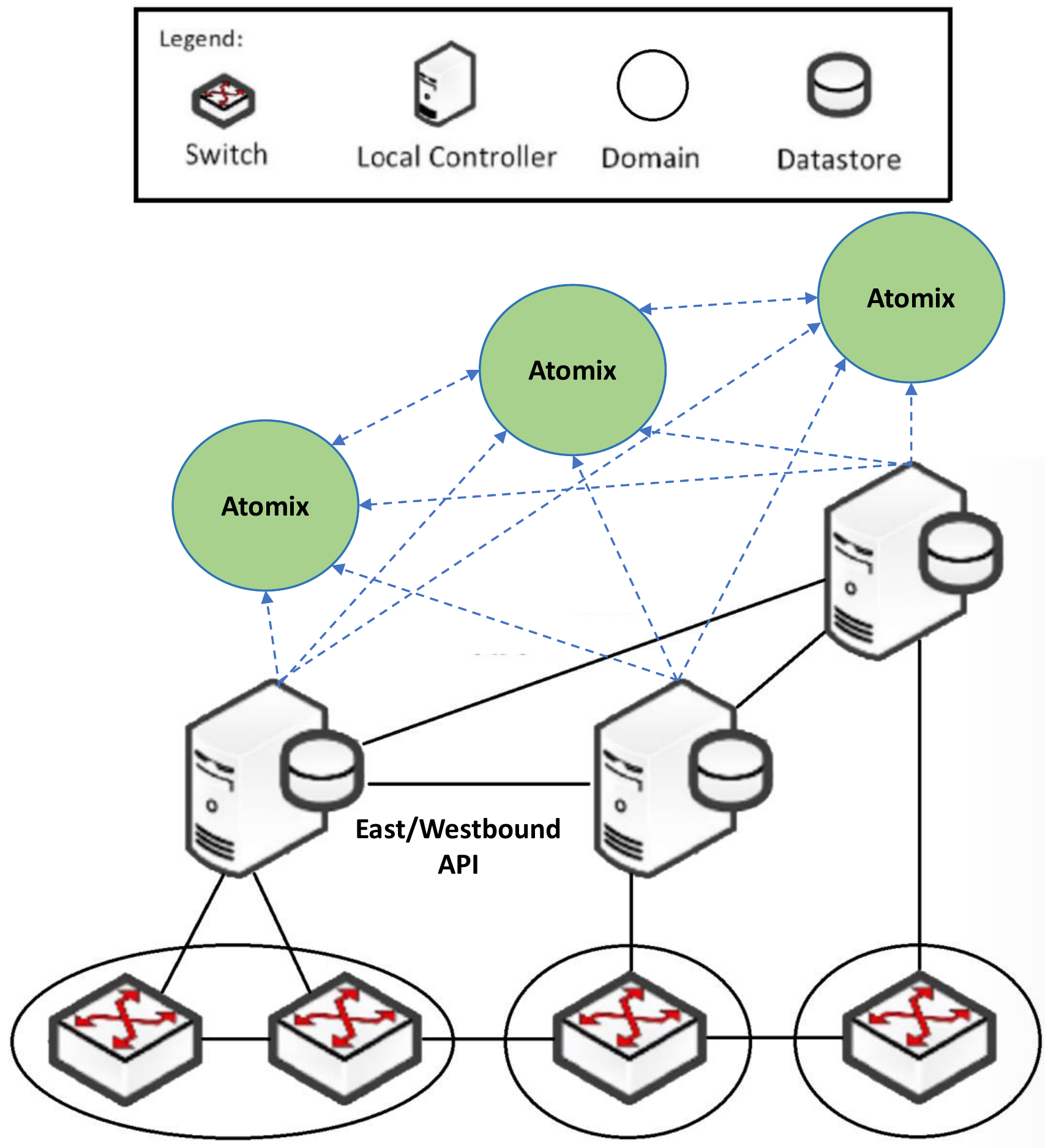} 

\caption{\centering{  Distributed, Flat SDN Controller Architecture \cite{oktian2017distributed}}}

\label{fig:SDN_arch}
\end{figure}

Figure~\ref{fig:SDN_arch} illustrates the distributed SDN architecture to manage the framework and which incorporates the aforementioned characteristics. The selected distributed flat SDN architecture divides the network horizontally into multiple areas or clusters, and each cluster is handled by a single controller. 
The SDN controllers in this architecture are interconnected and each controller has a global view of the network. 
ONOS can deploy distributed SDN controllers that utilize the flat architecture to achieve a logically distributed software-defined network~\cite{8187644}. 

To emulate and implement the D-SDN, first the Atomix and ONOS clusters are built using Docker containers in a local virtual machine (VM) with 6GB of RAM and 5 logical cores of AMD Ryzen 7 1800x installed on a local personal device using ONOS version 2.3, Openflow version 1.3, and Atomix version 3.1.5. 
Then, 118 hosts and 45 Open vSwitches using OpenFlow 1.3 were constructed using a Python script and Mininet 2.3 APIs. 
Open vSwitch %~\cite{openvswitch}
is an open-source distributed virtual multi-layer switch solution and one of the most widely used OpenFlow implementations.

The smart grid's communication layer is a three-controller SDN framework. Several approaches could be used, such as a hierarchical model or vertical architecture.
In this approach, a root controller manages the local controllers, and each local controller manages a network domain or cluster.
The root controller coordinates local controllers. To perform inter-domain functions, each local controller queries the root controller for global network state information, which adds latency and still provides a single point of failure~\cite{oktian2017distributed}.  To avoid a single point of failure, a flat, distributed topology is used where the three controllers share the same network authority. We also use ONOS to control packet flow in the smart grid system.
Unlike the hierarchical model, that does not connect local controllers, controllers in the flat model have direct connectivity, and every controller has equal access to the network. Therefore, ONOS can quickly redistribute controller load to optimize network performance. Furthermore, ONOS and Atomix can dynamically move workloads from a down controller to the remaining controllers to reduce single points of failure.
Atomix %~\cite{onos_wiki}
is a reactive Java framework for constructing scalable fault-tolerant distributed networks and manages our ONOS controllers as illustrated in Figure~\ref{fig:SDN_arch}. It is in charge of ONOS cluster management, service discovery, and data storage. 
Because of the Atomix framework, ONOS controllers may be rapidly discovered and removed if needed.

The design of a logically distributed framework with low latency and improved resilience of the communication layer of the IEEE-118 bus system for a smart grid is of interest. 
To incorporate these characteristics into the SDN framework, it was realized a need of a controller that could respond to network demands swiftly, provide global oversight to reroute traffic based on link status, and had a proven performance in both literature and industry. 
Given these factors, the ONOS controller was chosen. 

\section{Case Study}
\label{Case Study}

\begin{figure}[!h]
\centering
\includegraphics[width=.9\columnwidth]{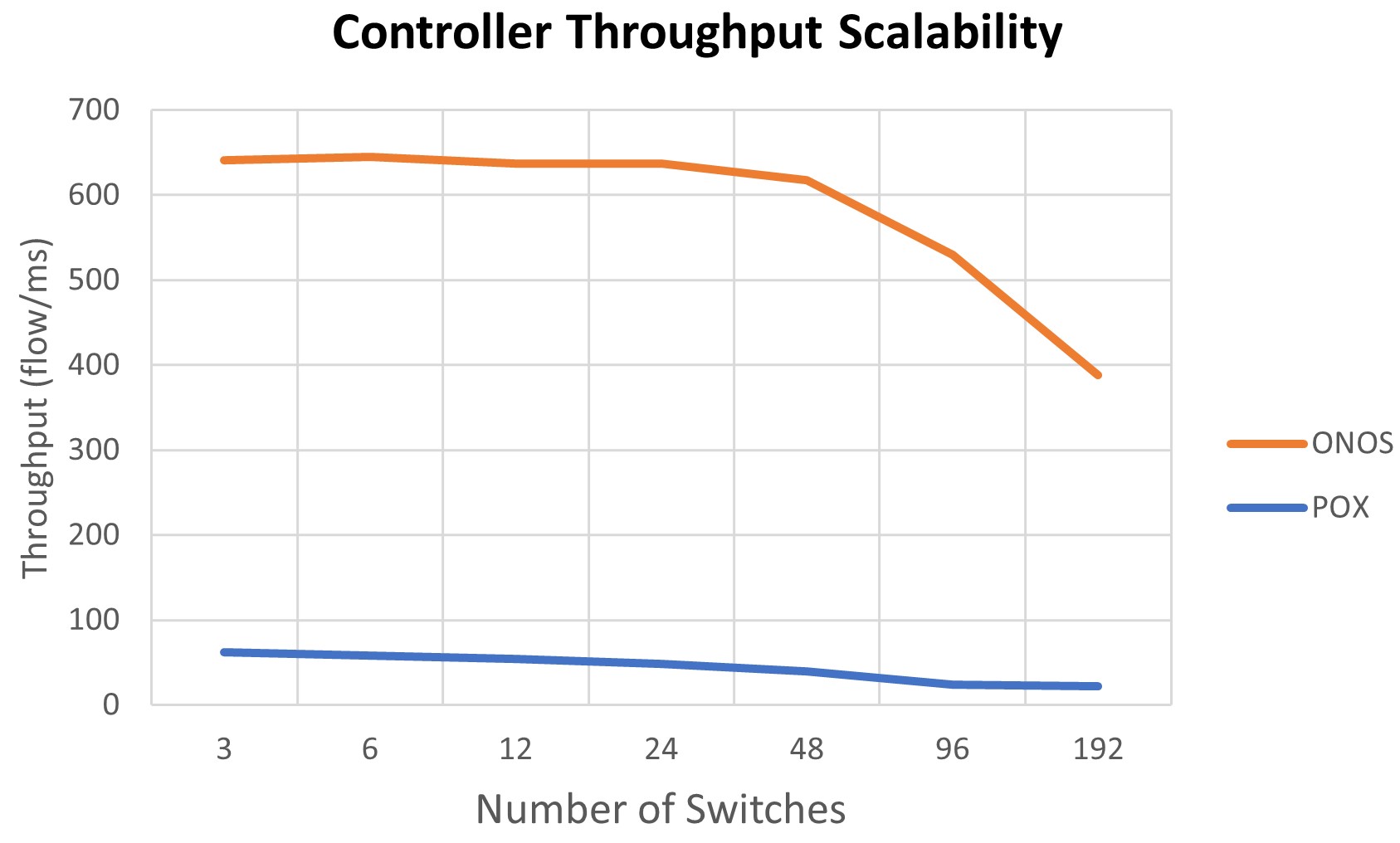}
\caption{\centering{ Throughput Scalability of Controllers}}
\label{fig:Scalability_Throughput_Benchmark}
\end{figure}

\begin{figure}[t!]
\centering
\includegraphics[width=.9\columnwidth]{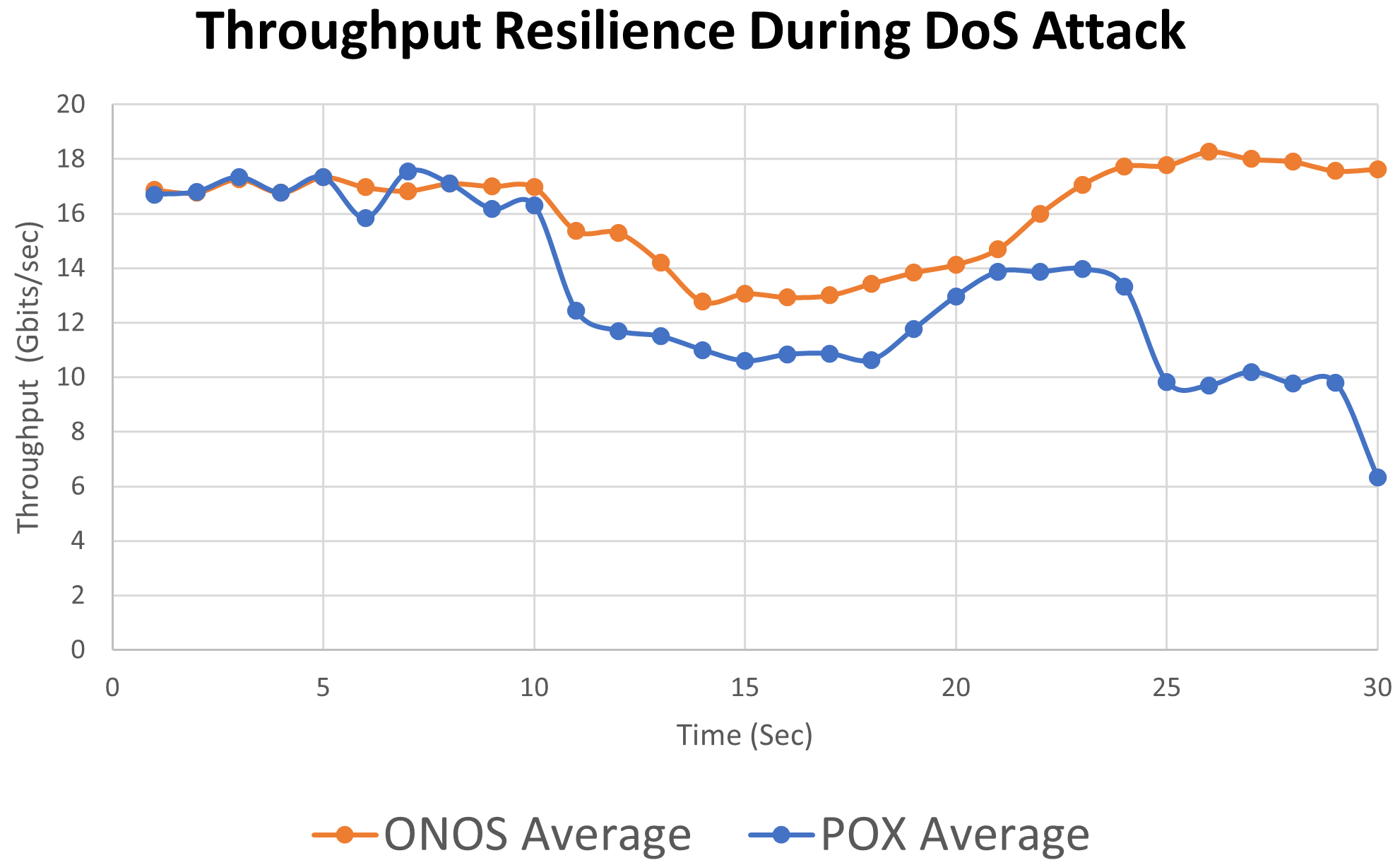} 
\caption{\centering{Throughput Resilience During DoS Attack}}
\label{fig:DoS_Throughput}
\end{figure}

A case study was performed to compare the performance of the proposed ONOS managed flat, three-controller, distributed SDN (D3-SDN), against the POX controller approach.
The results are presented in Figure~\ref{fig:Scalability_Throughput_Benchmark}, Table~\ref{tab:Latency}, and Figure~\ref{fig:DoS_Throughput}. 

First, to evaluate the scalability of D3-SDN we used CBench, a popular controller benchmark tool to test and record throughput performance. These results are shown in Figure~\ref{fig:Scalability_Throughput_Benchmark}. 
For this experiment, the  controller configurations ranged from 3 - 192 switches. 
As shown in Figure~\ref{fig:Scalability_Throughput_Benchmark}, the ONOS controllers have a higher initial throughput and maintain a higher throughput as the demand for more switches increases. 
As the number of switches that must be accounted for by each controller increases, the throughput performance of both controllers decreases. 
However, ONOS performs more effectively than POX overall.

Next, we compared the latency induced by each approach, based on the analysis in Section~\ref{sec:backgroundstats}. We determined the propagation delay, $D_{PD}$, for the emulated testbed by estimating the line lengths of the IEEE 118-bus system based on the per-unit to total reactance transformation presented in \cite{118buslength}. 
From the same reference we obtained each of the 117 line lengths based on area voltage level, after which an average line length was obtained. 
Then $D_{PD}$ was calculated  by dividing the average line length by the recorded link speed of $200,000$km per second \cite{coffey_2017} to obtain an average $D_{PD}$ of 203.307$\mu$s. We add variability by incorporating Gaussian noise to each $D_{PD}$ then assigning the value to each link's delay in our emulated testbed.

To measure the latency performance of our proposed ONOS D3-SDN approach and the traditional POX approach, we used the Sockperf % \cite{mellanox}
software to perform a ping pong test. A ping pong test is when computer A sends a packet to computer B (ping) requesting for computer B to send back an echo response (pong).
Sockperf operates by the client sending packets to the server, which then returns all or a portion of the packets to the client. 
It measures roundtrip time, $RTT$, between the two machines along a specific network path with packets of varying sizes. The latency for a particular one-way link between two computers is the $RTT$ divided by two. As our two nodes, we chose node 1 and node 112, the nodes with that required the most hops to communicate, as the Sockperf server and client, respectively and measured the latency performance on both $UDP$ and $TCP$ packets.
Table~\ref{tab:Latency} displays the results for average latency. 
The D3-SDN ONOS architecture reduces $UDP$ latency by $\sim24\%$ and $TCP$ latency by $\sim31\%$ over the POX controller approach, demonstrating that ONOS responds to network demands faster than POX.

\begin{table}[t!]
\centering
\renewcommand{\arraystretch}{1.1}
\centering
\caption{Average Latency Results for Legacy and ONOS Networks (UDP: User Datagram Protocol, TCP: Transmission Control Protocol).  }
\label{tab:Latency}
    \captionsetup{justification=centering}
\centering
\begin{tabular}{|c|c|c|}
\hline
\multirow{1}{*}{\textbf{ Network Type}} & {\textbf{Avg. UDP  Latency ({$\mu$s})}}
& {\textbf{Avg. TCP Latency ({$\mu$s})}}  \\
\hline 
        {ONOS}   &  28.727 & 28.846  \\
        \hline
        {POX}    & 37.876  & 42.345    \\
        \hline
\end{tabular}
\end{table}

Finally, to measure the  throughput resilience during and after a DoS attack for the D3-SDN ONOS approach versus the traditional POX approach, we conducted a DoS attack on the frameworks using iPerf3 and hping. 
%iPerf3 \cite{mortimer2018iperf3} and hping \cite{hping}
A server on node 1 and a client on node 2 are emulated using iPerf. 
To keep the comparison fair between controllers, for each attack simulation for both controllers, the client and server are instructed to have a transfer rate of 18 Gbits/sec for a simulation time of 30 seconds. 
After 10 seconds, the attacker, node 3, is instructed to "flood" the client, node 2, with approximately $80,000$ packets as fast as possible, with an IAT of $<100$ microseconds. 
After 20 seconds, the DoS attack is stopped, and the simulation continues for 10 additional seconds. 
This attack scenario is repeated three times for each controller configuration and the results were recorded. 

Figure~\ref{fig:DoS_Throughput} shows the results. 
For the first 10 seconds, there is a natural initial decrease and variability in network throughput due to the controllers spending the necessary time to learn the network routes. At 10 seconds, the attack begins and both controllers experience throughput decreases as the attacker begins to attack the client. This is also expected behavior. During the DoS attack, however, the POX controller lost 41.11\% percent of its throughput, while D3-SDN ONOS lost only 29.77\% percent. Therefore, during our DoS attack scenario, ONOS retains an average $\sim11\%$ greater throughput than POX.

After the attack, the POX controller experiences the worst throughput from having been overwhelmed with the increased DoS demand. At its lowest point of the throughput measurements, the throughput drops to $\sim 6$  Gbits/s, which is a throughput loss of throughput of $\sim 66\%$  compared to the virtual $\sim 0\%$ lost experienced by D3-SDN ONOS approach at the time of simulation completion. This demonstrates the D3-SDN ONOS approach's resilience and its capacity to handle increased network demands, albeit fraudulent in the form of a DoS attack in this instance. Furthermore,  ONOS is able to recover quickly after these attack instances, which is advantageous when developing robust communication layers for smart grids.

Since POX fails in terms of maintaining throughput and the D3-SDN ONOS approach accommodates the DoS traffic, it is important to be able to determine that an attack has occurred and identify which type. In previous work \cite{Allen2022Starke}, an algorithm called cross-layer ensemble corrdet is presented for cyber threat detection. Its performance is proven to exceed state-of-the-art physics-based and machine learning-based methodologies. 
 In our previous study~\cite{aljohanicross}, the algorithm's pseudo code and implementation are described. Each row represents a point in time for the relevant measurements, whereas each column represents a grid measurement point. A a 691 x 10,000
matrix of data points is built in Mininet, where one row of data is considered a sample of measurements for the entire grid, and
Mininet creates one data point in around 4 seconds, or 45 minutes for each network sample, or row of data. 
The ML model requires 10,000 rows of data, which poses a time restriction. 
To address this limitation, SimComponents, a network simulator tool, was used to reduce the amount of time spent obtaining the network data. The SimPy and component toolkit SimComponents are utilized in the communication layer in order to simulate the network traffic one would encounter in the ONOS-managed D-SDN. SimPy is a Python-based discrete-event simulation toolbox. 
Active components including packets, packet generators, packet sinks, switch ports, and port monitors may be simulated with this tool. The SimComponents toolkit is used to define and replicate various components and their features. SimComponent creates the data required for one data sample in around $0.80$ seconds, which is significantly faster than Mininet, allowing the completion of the necessary dataset to acquire the results for the CECD-AS algorithm technique, as shown in Table~\ref{tab:CECD}.

The real-time CECD-AS algorithm works exceedingly well for a range of cyber threats mentioned in this research, as seen in Table~\ref{tab:CECD}. The improved detection is the result of the integration (combined) data from communication and power grid. 

\begin{table}[h]
\centering
\centering
\caption{Performance results for FDI, DoS, and MITM attacks (FDI: False Data Injection attacks, DoS: Denial of Service attacks, MITM: Man In The Middle attacks)}
\label{tab:CECD}
    \captionsetup{justification=centering}

\resizebox{1\columnwidth}{!}{%  
\begin{tabular}{@{}|c|c|c|c|c|@{}}

\hline
\multirow{2}{*}{\textbf{Attack type}} & {\textbf{Accuracy}}  & {\textbf{Precision}} & {\textbf{Recall}} & {\textbf{F1-score}} \\
      
& \textbf{$\mu_{cv}$} $\pm$ \textbf{$\sigma_{cv}$} & \textbf{$\mu_{cv}$} $\pm$ \textbf{$\sigma_{cv}$} & \textbf{$\mu_{cv}$} $\pm$ \textbf{$\sigma_{cv}$} & \textbf{$\mu_{cv}$} $\pm$ \textbf{$\sigma_{cv}$} \\% 
\hline
        MITM & 92.48 $\pm$ 00.20 & 91.65 $\pm$ 00.29 & 86.41 $\pm$ 00.28 & \textbf{88.91 $\pm$ 00.24} \\ %\midrule
        \hline
        {FDI} & 99.95 $\pm$ 00.01 & 99.46 $\pm$ 00.34 & 99.87 $\pm$ 00.13 & \textbf{99.61 $\pm$ 00.17} \\
        %\midrule%%%%
        \hline
        
        {DoS} & 99.88 $\pm$ 00.07 & 99.75 $\pm$ 00.09 & 99.80 $\pm$ 00.16 & \textbf{99.78 $\pm$ 00.08} \\
        %\midrule
        \hline
        {FDI-DoS} & 99.63 $\pm$ 00.08 & 98.42 $\pm$ 00.26 & 99.95 $\pm$ 00.04 & \textbf{99.20 $\pm$ 00.15} \\

        \hline

\end{tabular}%
}
\hspace{1cm}

\end{table}

\section{Conclusion}
\label{Conclusion}
SDN enables users to manage smart grid communications with greater visibility, control, and reactivity. Deploying a set of ONOS clusters with equal permissions reduces the possibility of a single point of failure that may arise when using a single controller.  This study evaluates a distributed three ONOS controller managed SDN architecture for smart grids and compares it with the common alternative POX controller. 
The ONOS managed architecture significantly outperformed the POX controller framework by reducing latency by $>20\%$ and increasing controller throughput ten-fold, resulting in increased scalability and significant throughput resiliency during DoS attacks. 
These performance results demonstrate that the D3-SDN is the better option over POX implementations regarding cyber security of the smart grid.

\bibliographystyle{IEEEtran}
\bibliography{conference_101719}

\begin{thebibliography}{10}
\providecommand{\url}[1]{#1}
\csname url@samestyle\endcsname
\providecommand{\newblock}{\relax}
\providecommand{\bibinfo}[2]{#2}
\providecommand{\BIBentrySTDinterwordspacing}{\spaceskip=0pt\relax}
\providecommand{\BIBentryALTinterwordstretchfactor}{4}
\providecommand{\BIBentryALTinterwordspacing}{\spaceskip=\fontdimen2\font plus
\BIBentryALTinterwordstretchfactor\fontdimen3\font minus
  \fontdimen4\font\relax}
\providecommand{\BIBforeignlanguage}[2]{{%
\expandafter\ifx\csname l@#1\endcsname\relax
\typeout{** WARNING: IEEEtran.bst: No hyphenation pattern has been}%
\typeout{** loaded for the language `#1'. Using the pattern for}%
\typeout{** the default language instead.}%
\else
\language=\csname l@#1\endcsname
\fi
#2}}
\providecommand{\BIBdecl}{\relax}
\BIBdecl

\bibitem{monticelli1999state}
A.~Bretas, N.~Bretas, J.~London, and B.~Carvalho, \emph{Cyber-Physical Power
  Systems State Estimation}.\hskip 1em plus 0.5em minus 0.4em\relax Elsevier,
  2021, vol.~1.

\bibitem{Allen2022Starke}
\BIBentryALTinterwordspacing
A.~Starke, K.~Nagaraj, C.~Ruben, N.~Aljohani, S.~Zou, A.~Bretas, J.~McNair, and
  A.~Zare, ``Cross-layered distributed data-driven framework for enhanced smart
  grid cyber-physical security,'' \emph{IET Smart Grid}, vol. n/a, no. n/a.
  [Online]. Available:
  \url{https://ietresearch.onlinelibrary.wiley.com/doi/pdf/10.1049/stg2.12070}
\BIBentrySTDinterwordspacing

\bibitem{agnew2022implementation}
D.~Agnew, N.~Aljohani, R.~Mathieu, S.~Boamah, K.~Nagaraj, J.~McNair, and
  A.~Bretas, ``Implementation aspects of smart grids cyber-security
  cross-layered framework for critical infrastructure operation,''
  \emph{Applied Sciences}, vol.~12, no.~14, p. 6868, 2022.

\bibitem{aljohanicross}
N.~Aljohani, D.~Agnew, K.~Nagaraj, S.~A. Boamah, R.~Mathieu, A.~S. Bretas,
  J.~McNair, and A.~Zare, ``Cross-layered cyber-physical power system state
  estimation towards a secure grid operation,'' in \emph{2022 IEEE Power \&
  Energy Society General Meeting (PESGM)}.\hskip 1em plus 0.5em minus
  0.4em\relax IEEE, 2022, pp. 1--5.

\bibitem{starke2018toward}
A.~Starke, J.~McNair, R.~Trevizan, A.~Bretas, J.~Peeples, and A.~Zare, ``Toward
  resilient smart grid communications using distributed sdn with ml-based
  anomaly detection,'' in \emph{International Conference on Wired/Wireless
  Internet Communication}.\hskip 1em plus 0.5em minus 0.4em\relax Springer,
  2018, pp. 83--94.

\bibitem{kaur2014network}
S.~Kaur, J.~Singh, and N.~S. Ghumman, ``Network programmability using pox
  controller,'' in \emph{ICCCS International conference on communication,
  computing \& systems, IEEE}, vol. 138.\hskip 1em plus 0.5em minus 0.4em\relax
  sn, 2014, p.~70.

\bibitem{cokic2019software}
M.~Cokic and I.~Seskar, ``Software defined network management for dynamic smart
  grid traffic,'' \emph{Future Generation Computer Systems}, vol.~96, pp.
  270--282, 2019.

\bibitem{mahmood2021s}
H.~Mahmood, D.~Mahmood, Q.~Shaheen, R.~Akhtar, and W.~Changda, ``S-dps: An
  sdn-based ddos protection system for smart grids,'' \emph{Security and
  Communication Networks}, vol. 2021, 2021.

\bibitem{ghosh2017security}
U.~Ghosh, P.~Chatterjee, and S.~Shetty, ``A security framework for sdn-enabled
  smart power grids,'' in \emph{2017 IEEE 37th International Conference on
  Distributed Computing Systems Workshops (ICDCSW)}.\hskip 1em plus 0.5em minus
  0.4em\relax IEEE, 2017, pp. 113--118.

\bibitem{qureshi2020distributed}
K.~N. Qureshi, R.~Hussain, and G.~Jeon, ``A distributed software defined
  networking model to improve the scalability and quality of services for
  flexible green energy internet for smart grid systems,'' \emph{Computers \&
  Electrical Engineering}, vol.~84, p. 106634, 2020.

\bibitem{hussain2019model}
R.~Hussain and M.~U. Bashir, ``Model to improve scalability and quality of
  services in software define networking,'' in \emph{2019 2nd International
  Conference on Communication, Computing and Digital systems (C-CODE)}.\hskip
  1em plus 0.5em minus 0.4em\relax IEEE, 2019, pp. 28--33.

\bibitem{kreutz2014software}
D.~Kreutz, F.~M. Ramos, P.~E. Verissimo, C.~E. Rothenberg, S.~Azodolmolky, and
  S.~Uhlig, ``Software-defined networking: A comprehensive survey,''
  \emph{Proceedings of the IEEE}, vol. 103, no.~1, pp. 14--76, 2014.

\bibitem{sun2020detecting}
S.~Sun, X.~Fu, B.~Luo, and X.~Du, ``Detecting and mitigating arp attacks in
  sdn-based cloud environment,'' in \emph{IEEE INFOCOM 2020-IEEE Conference on
  Computer Communications Workshops (INFOCOM WKSHPS)}.\hskip 1em plus 0.5em
  minus 0.4em\relax IEEE, 2020, pp. 659--664.

\bibitem{mamushiane2021qos}
L.~Mamushiane and T.~Shozi, ``A qos-based evaluation of sdn controllers: Onos
  and opendaylight,'' in \emph{2021 IST-Africa Conference (IST-Africa)}.\hskip
  1em plus 0.5em minus 0.4em\relax IEEE, 2021, pp. 1--10.

\bibitem{github}
\BIBentryALTinterwordspacing
``Mininet/mininet: Emulator for rapid prototyping of software defined
  networks.'' [Online]. Available:
  \url{https://github.com/mininet/mininet#readme}
\BIBentrySTDinterwordspacing

\bibitem{jain1986packet}
R.~Jain and S.~Routhier, ``Packet trains--measurements and a new model for
  computer network traffic,'' \emph{IEEE journal on selected areas in
  Communications}, vol.~4, no.~6, pp. 986--995, 1986.

\bibitem{haviv2009queues}
M.~Haviv, ``Queues--a course in queueing theory,'' \emph{The Hebrew University,
  Jerusalem}, vol. 91905, 2009.

\bibitem{8187644}
F.~Bannour, S.~Souihi, and A.~Mellouk, ``Distributed sdn control: Survey,
  taxonomy, and challenges,'' \emph{IEEE Communications Surveys \& Tutorials},
  vol.~20, no.~1, pp. 333--354, 2018.

\bibitem{oktian2017distributed}
Y.~E. Oktian, S.~Lee, H.~Lee, and J.~Lam, ``Distributed sdn controller system:
  A survey on design choice,'' \emph{computer networks}, vol. 121, pp.
  100--111, 2017.

\bibitem{118buslength}
\BIBentryALTinterwordspacing
PSCAD. (2018) Ieee 118 bus system. [Online]. Available:
  \url{https://www.pscad.com/knowledge-base/download/ieee_118_bus_technical_note.pdf}
\BIBentrySTDinterwordspacing

\bibitem{coffey_2017}
J.~Coffey, \emph{Latency in optical fiber systems}, 2017.

\end{thebibliography}

\end{document}